\documentclass[twocolumn,amsmath,amssymb,prb]{revtex4}

\usepackage{graphicx,subfigure}
\usepackage{dcolumn}
\usepackage{bm}
\usepackage{color}
\usepackage{ulem}

\newcommand{\irte}{IrTe$_2$}

\newif\ifCMbool
\CMbooltrue

\ifCMbool
\newcommand{\CMs}[1]{\textcolor{red}{{ \sout{#1} }}}
\else
\newcommand{\CMs}[1]{\textcolor{red}{{ }}}
\fi

\begin{document}

\preprint{APS/123-QED}

\title{Robustness of the charge-ordered phases in IrTe$_2$ against photoexcitation}

\author{C. Monney$^{1,2,*}$, A. Schuler$^{1}$, T. Jaouen$^{2}$, M.-L. Mottas$^{2}$, Th. Wolf$^{3}$, M. Merz$^{3}$, M. Muntwiler$^{4}$, L. Castiglioni$^{1}$, P. Aebi$^{2}$, F. Weber$^{3}$, M. Hengsberger$^{1}$}

\affiliation{
$^1$Physik-Institut, Universit\"at Z\"urich, Winterthurerstrasse 190, 8057 Z\"urich, Switzerland,\\
$^2$D\'epartement de Physique and Fribourg Center for Nanomaterials, Universit\'e de Fribourg, 1700 Fribourg, Switzerland,\\
$^3$Institute of Solid State Physics, Karlsruhe Institute of Technology, 76131 Karlsruhe, Germany,\\
$^4$Paul Scherrer Institut, 5232 Villigen PSI, Switzerland\\
$^*$claude.monney@unifr.ch
}

\date{\today}

\begin{abstract} 
We present a time-resolved angle-resolved photoelectron spectroscopy study of \irte, which undergoes two first-order structural and charge-ordered phase transitions on cooling below 270 K and below 180 K.
The possibility of inducing a phase transition by photoexcitation with near-infrared femtosecond pulses is investigated in the charge-ordered phases. We observe changes of the spectral function occuring within a few hundreds of femtoseconds and persisting up to several picoseconds, which we interpret as a partial photoinduced phase transition (PIPT). The necessary time for photoinducing these spectral changes increases with increasing photoexcitation density and reaches timescales longer than the rise time of the transient electronic temperature. We conclude that the PIPT is driven by a transient increase of the lattice temperature following the energy transfer from the electrons. However, the photoinduced changes of the spectral function are small, which indicates that the low temperature phase is particularly robust against photoexcitation. We suggest that the system might be trapped in an out-of-equilibrium state, for which only a partial structural transition is achieved.
\end{abstract}

\maketitle

\section{Introduction}

Transition metal dichalcogenides (TMDCs) are layered quasi-two-dimensional materials which provoked tremendous attention in the last years, because of the possibility of easily reducing their size down to the monolayer and of their interesting properties despite their relative chemical simplicity. This makes them attractive candidates for modern devices.
TMDCs were intensively studied in the 1970s as they are hosting charge density wave (CDW) phases at low temperature, involving a phase transition which is usually enabled by an enhancement in the electronic susceptibility due to nesting of the low-dimensional Fermi surface. 
Furthermore, recent research on materials with 4$d$ and 5$d$ electrons has highlighted the importance of strong spin-orbit coupling and a significant Hund's coupling for the physics of correlated materials\cite{HundRev,SutterNatComm}. In this framework, \irte\ with 5$d$ valence electrons in dispersing valence states and showing charge ordering at low temperature is intriguing and raises the question of whether Mott physics develops at low temperature\cite{KoNatureComm}. It offers the interesting case of a TMDC with potentially strong electronic correlations in spin-orbit coupled bands.

\irte\ undergoes a first-order structural phase transition at about $T_{c1}=270$ K to a phase with a charge-ordered state characterized by a wave vector $q_1=(1/5,0,1/5)$ with respect to the room temperature unit cell vectors. The first-order phase transition is accompanied by a change of its unit cell symmetry from trigonal to monoclinic\cite{MatsumotoEarly}.
It involves a large jump in transport and magnetic properties\cite{MatsumotoEarly} as well as in heat capacity\cite{Fang2013}. We call this phase LT1. This phase transition displays also hysteretic magnetic and electrical behaviors. At about $T_{c2}=180$ K, a second phase transition occurs and the charge ordering wave vector changes to $q_2=(1/8,0,1/8)$. We call it LT2. These phases with different charge ordering patterns have stimulated many scanning tunneling microscopy studies which evidenced the occurence of additional ordering patterns at the surface\cite{KoNatureComm,ChenSTM,DaiSTM,HsuSTM,KimSTM,LiSTM,BodeSTM}.
In parallel, motivated by the TMDC structure of \irte\ and the possibility of finding a new CDW phase, many angle-resolved photoelectron spectroscopy (ARPES) investigations have targeted this material \cite{KoNatureComm,KimARPES,OotsukiJPhys,OotsukiJPS,OotsukiJPSJ,QianARPES}. All of these ARPES studies have provided evidence for spectral changes in the low-energy electronic structure down to 3 eV below the Fermi level $E_F$. These changes appear abruptly across $T_{c1}$, while more subtle differences occur across $T_{c2}$. However, no CDW gap has been observed, nor any significant possibility for nesting of the Fermi surface, although the inner bands (near $\bar{\Gamma}$) disappear below $T_{c1}$ and might therefore be gapped. Interestingly, a new band has been found in the LT1 phase at about 0.5 eV below $E_F$\cite{KoNatureComm,QianARPES}.
The complexity of the band structure of \irte, together with its rich phase diagram involving many ordering patterns, makes it difficult to determine the mechanism behind its phase transitions. 
It has been put forward that the lattice reconstruction induces the lifting of an orbital degeneracy which stabilizes the LT1 phase. This can be viewed as a Jahn-Teller effect\cite{OotsukiJPhys,KimARPES}. Another study proposed a Mott phase involving spin-orbit coupled states\cite{KoNatureComm} to be at the origin of the phase transitions in \irte. In this framework, time-resolved techniques can provide new information about the nature of these phase transitions by discriminating such different mechanisms on the femtosecond to picosend timescales. The suppression of the Mott phase in TaS$_2$ has been shown to be as fast as the photoexcitation pulse\cite{PetersenTaS2,PerfettiTaS2,HellmanNatComm}, while the suppression of phases involving a structural reconstruction requires often longer times\cite{HellmanNatComm,FriggeNature,SchmittTbTe}.

Here we perform a systematic time-resolved ARPES study of \irte\ in its low-temperature charge-ordered phases LT1 and LT2 and investigate the possibility of photoinducing the phase transition to the room temperature (RT) phase. Using 6 eV photons, we single out a specific band in its electronic structure, which we relate to these charge-ordered states. After photoexciting the material, a small transient shift of this band is observed and interpreted as the result of a partial PIPT. It is shown that the necessary time for photoinducing this spectral change increases with increasing photoexcitation density. The timescale of this spectral change turns out to be even slower than the rise time of the transient electronic temperature. We conclude that this partial PIPT is driven by the transient increase of the lattice temperature, after the energy transfer from the electrons.
However, despite the high photoexcitation densities used here, the photoinduced changes are small, indicating that \irte\ is very robust against photoexcitation with near-infrared pulses. We propose that the system is not transiting to the RT phase, but is trapped in an out-of-equilibrium state, for which only a partial structural transition is achieved.

\section{Experimental details}

\begin{figure*}
\begin{center}
\includegraphics[width=17cm]{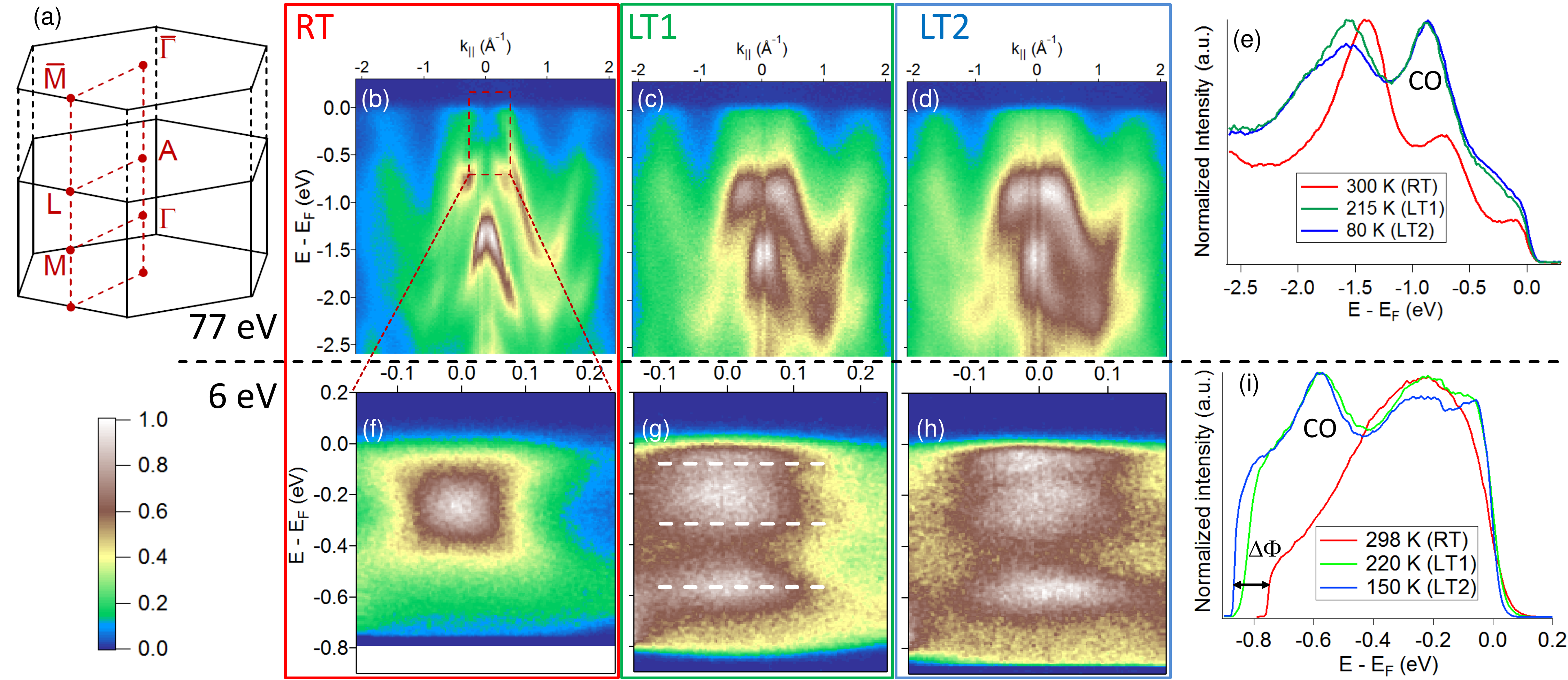}
\end{center}
\caption{\label{fig_1}
(a) Bulk and surface Brillouin Zone of \irte\ in its room temperature phase.
(b,c,d) Static ARPES data acquired with 77 eV photons along $\bar{\Gamma}\bar{M}$ at (b)	300 K, (c) 215 K and (d) 80 K, corresponding to the RT, LT1 and LT2 phases, respectively. (e) Angle-integrated energy distribution curves (normalized to their maximum) integrated $\pm 0.3$ \AA$^{-1}$ around $\bar{\Gamma}$ in the data of graphs (b,c,d).
Static ARPES maps acquired with 6 eV photons along $\bar{\Gamma}\bar{M}$ at (f)	298 K, (g) 230 K and (h) 160 K. The white dashed lines in graph (g) indicate the three flat bands mentioned in the main text. (i) Angle-integrated energy distribution curves (normalized to their maximum) from the data of graphs (f,g,h). A significant shift of the work function $\Delta\Phi\cong0.15$ eV is observed between the RT and LT1/LT2 phases. Note that the ARPES maps in graphs (f,g) are not centered at $\bar{\Gamma}$.
}
\end{figure*}

Single crystals of \irte\ were grown using the self-flux method. Resulting single crystals were characterized by magnetization measurements and single crystal x-ray diffraction studies, which confirms that $T_{c1}=270$ K and $T_{c2}= 180$ K. The static ARPES data were acquired at the PEARL beamline\cite{PEARL} of the Swiss Light Source at the Paul Scherrer Institute (Switzerland).At 77 eV photon energy, the total energy resolution was 60 meV and the momentum resolution was 0.03 \AA$^{-1}$.
The time-resolved ARPES data were obtained using light pulses produced by means of a commercial femtosecond oscillator (Coherent Mira Seed) and amplified in a high repetition rate (30 kHz) regenerative pulse amplifier (RegA 9050). \irte\ samples were cleaved in a base pressure of $10^{-10}$ mbar and normal emission of photoelectrons was lying along $(0001)$. The samples were excited by $p$-polarized laser pulses at 825 nm (1.5 eV). The samples were probed by $p$-polarized laser pulses of 6.0 eV photons, generated by frequency-doubling the fundamental 825 nm pulses twice in $\beta-$barium borate crystals. At 6.0 eV photon energy, the total energy resolution was 50 meV, the momentum resolution was 0.01 \AA$^{-1}$ and the time cross-correlation was 250 fs (from the same sample). The absorbed fluences (absorbed excitation energy per unit area) appearing in this study are obtained by considering the measured incident fluences and a sample reflectivity\cite{Fang2013} of 50$\%$ at 825~nm. 
During the measurements with 6 eV photons, a bias voltage of $-3$ V with respect to ground in order to avoid the energy range of very low kinetic energies where the transmission of the electron detector almost vanishes. Furthermore this bias voltage helps to obtain sharper secondary electron edges for the determination of the sample work function. 
Sample temperature was measured with an accuracy of about $\pm$10 K.

\section{Results}

In Fig. \ref{fig_1}, we show static ARPES data of \irte\ taken with high and low photon energies at three different temperatures corresponding to the three different phases of \irte. With a photon energy of 77 eV, the full Brillouin zone (displayed in Fig. \ref{fig_1}(a)) is easily accessible with ARPES and at normal emission photoemission probes the region near $\Gamma$ (the wave vector $k_z$ perpendicular to the surface corresponds to about $4.1c^*$ with an inner potential\cite{KoNatureComm} of 13 eV, $c^*$ being the Brillouin zone size perpendicular to the surface). From now on, we adopt the surface Brillouin zone notation as marked by bars. At room temperature, we observe a few sharp and dispersive bands along the $\bar{\Gamma}\bar{M}$ direction, as shown in Fig. \ref{fig_1} (b). After cooling the sample to 215 K (Fig. \ref{fig_1} (c)), in the LT1 phase, the bands become suddenly broad in agreement with other ARPES studies\cite{KoNatureComm,OotsukiJPhys,OotsukiJPS,OotsukiJPSJ,QianARPES}, so that it is difficult to distinguish the different contributions to the low energy electronic structure of \irte. In Fig. \ref{fig_1} (d), we show ARPES data measured at 80 K in LT2, which are similar to what has been measured at 215 K (Fig. \ref{fig_1} (c)). Energy distribution curves (EDCs) integrated over $\pm 0.3$ \AA$^{-1}$ around $\bar{\Gamma}$ are shown in Fig. \ref{fig_1} (e) for these three temperatures. It clearly demonstrates the strong spectral changes occuring in photoemission after cooling through the first phase transition. Notice especially the peak at about $E-E_F=-0.9$ eV (labelled CO) which gains intensity in comparison to the peak at about -1.5~eV. Only small changes are visible in the spectra taken across the LT1-LT2 phase transition.

On the lower graphs of Fig. \ref{fig_1}, we show the static ARPES data taken with a photon energy of 6 eV. With such an excitation energy, the photoelectrons have typical kinetic energies $< 1$ eV, meaning that only the center of the surface Brillouin zone is probed, with an excellent momentum resolution. 
Globally, the photoemission intensity distribution over the whole ARPES maps is significantly different than what is measured with 77~eV photons. This is expected because of the different position of the probed initial states along the $k_z$-direction and the electronic structure of \irte\ displaying a significant dispersion along $k_z$, despite its atomic structure being similar to that of many TMDCs\cite{Fang2013}. 
In Fig. \ref{fig_1} (f), room temperature ARPES data indicate two bands dispersing close to each to other at the zone center. In comparison to Fig. \ref{fig_1} (b) (obtained with 77~eV photons), only the zone outlined by the red dashed line is probed. By comparing to the data of Ootsuki {\it et al.}, it appears that 6 eV photons probe the vicinity of the $A$ point along the $k_z$ direction \cite{OotsukiJPSJ}.
In addition, the final states of photoemission with 6 eV photons are expected to be significantly deviating from free electron final states\cite{HengsbergerCu,Miller6eV}, which might affect strongly the photoemission matrix elements, and thus the photoemission intensity distribution.
At 220~K in the LT1 phase, the ARPES spectrum is completely different, see Fig. \ref{fig_1} (g). One can now distinguish three flat bands (see dashed lines in graph (g)) instead of the two dispersing ones. When further decreasing the temperature to 150~K into the LT2 phase (Fig. \ref{fig_1} (h)), there are little changes. Fig. \ref{fig_1} (i) shows the comparison of the angle-integrated EDCs for these three temperatures for the data obtained with 6 eV photons. This highlights a few important differences: (i) In addition to the drastic changes between room temperature and low temperature, there is a significant difference in intensity for the peak at -0.6~eV relative to the peak at about -0.2~eV between LT1 and LT2. From now on, we will consider the peak at -0.6~eV as the spectral signature of the LT phases in \irte\ and we will name it the charge order (CO) peak for convenience. (ii) The work function is different in the three phases: 5.3~eV at RT, 5.15~eV in LT1 and 5.1~eV in LT2. We attribute these different values to the changes in the electronic and atomic structure of \irte\ in the charge-ordered phases LT1 and LT2.

\begin{figure}
\begin{center}
\includegraphics[width=8.5cm]{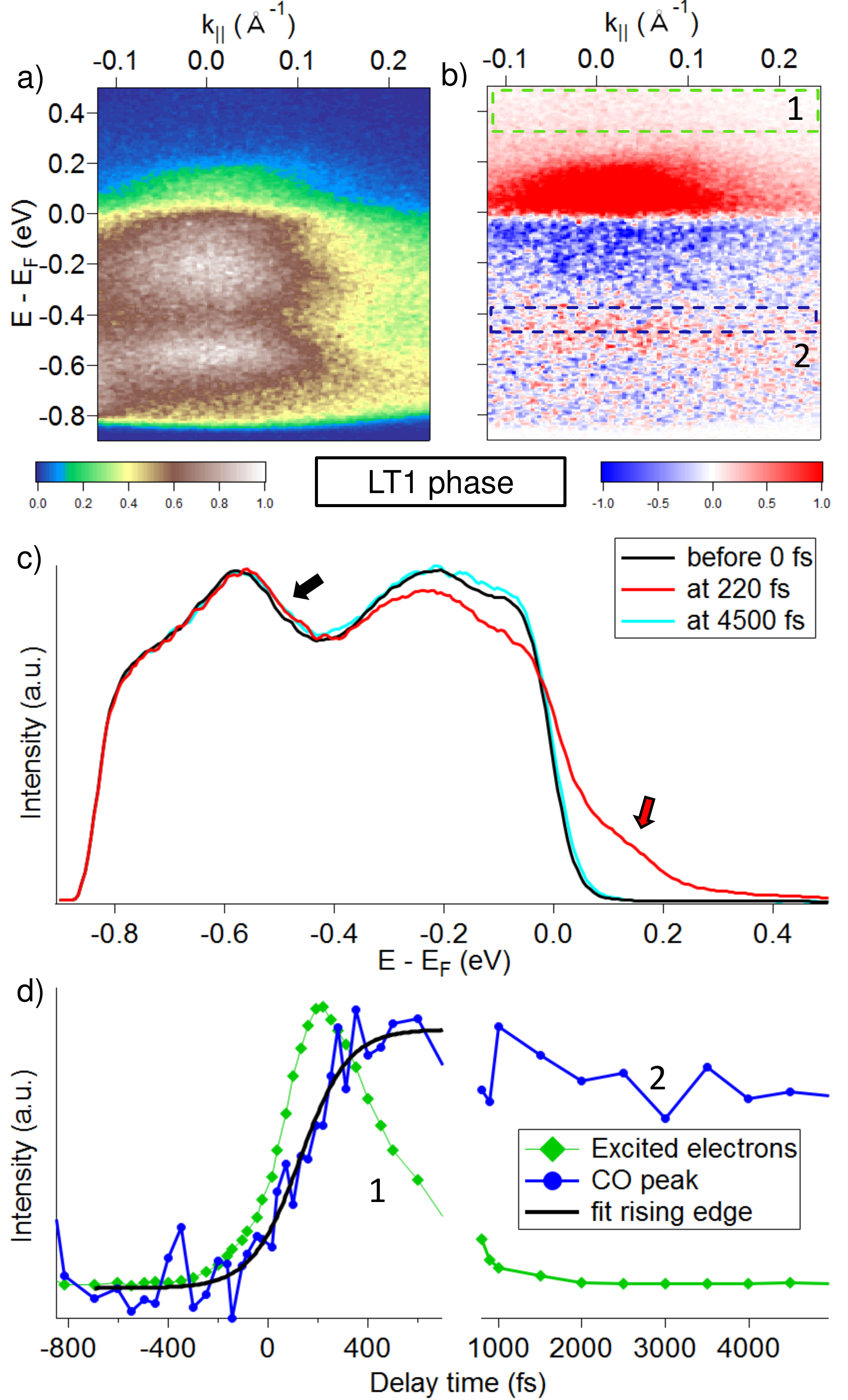}
\end{center}
\caption{\label{fig_2}
Time-resolved ARPES data in LT1 phase near $\bar{\Gamma}$, at 235 K. (a) Raw ARPES intensity map taken at + 220 fs and (b) the corresponding difference map (photoemission intensity difference between averaged data taken before 0 and data taken at 220 fs) for an absorbed fluence of 1.9 mJ/cm$^2$. (c) Corresponding (angle-integrated) EDCs at different delay times. (d) Transient intensities integrated in the box 1 (excited electrons) and 2 (CO peak) of graph (b), together with a step-like fit.
}
\end{figure}

Having established the spectral changes in static ARPES with 6 eV probe photons, we now move on to the time-resolved data. For this purpose, the samples are photoexcited with 1.5 eV photons. Given the complex electronic structure of \irte\ and the absence of a large band gap at low temperature, such an optical excitation can occur a priori via many direct transitions. In Fig. \ref{fig_2} (a), we show time-resolved ARPES data taken at 220 fs for \irte\ in the LT1 phase at 230 K for an absorbed excitation fluence of 1.9 mJ/cm$^2$. To emphasize the dynamical changes induced by the photoexcitation, we subtract from these ARPES data taken at 220 fs an average of the data taken at delays before time 0. This difference map at 220~fs is shown in Fig. \ref{fig_2} (b). It highlights the momentum-resolved changes in the photoemission intensity occuring at different energies. Excited electrons transiently populate states above $E_F$ (positive intensity change) and intensity is lost by the appearance of holes (negative intensity change) just below $E_F$. The excited electron distribution follows a non-trivial distribution with a shoulder at about 0.12~eV. This is better seen in the transient EDCs shown on Fig. \ref{fig_2} (c) for different times. At 220~fs, one distinguishes a shoulder above $E_F$ which is transiently populated and gives evidence for a band just above $E_F$ (see red arrow in \ref{fig_2} (c)). In addition to this population dynamics and corresponding hole dynamics, other transient changes occur at higher binding energies in the occupied states, below $-0.3$ eV. Interestingly, it affects also the CO peak, the spectral signature of the LT phases, which shifts towards $E_F$ (see the black arrow in the inset in Fig. \ref{fig_2} (c)).

\begin{figure}
\begin{center}
\includegraphics[width=8.5cm]{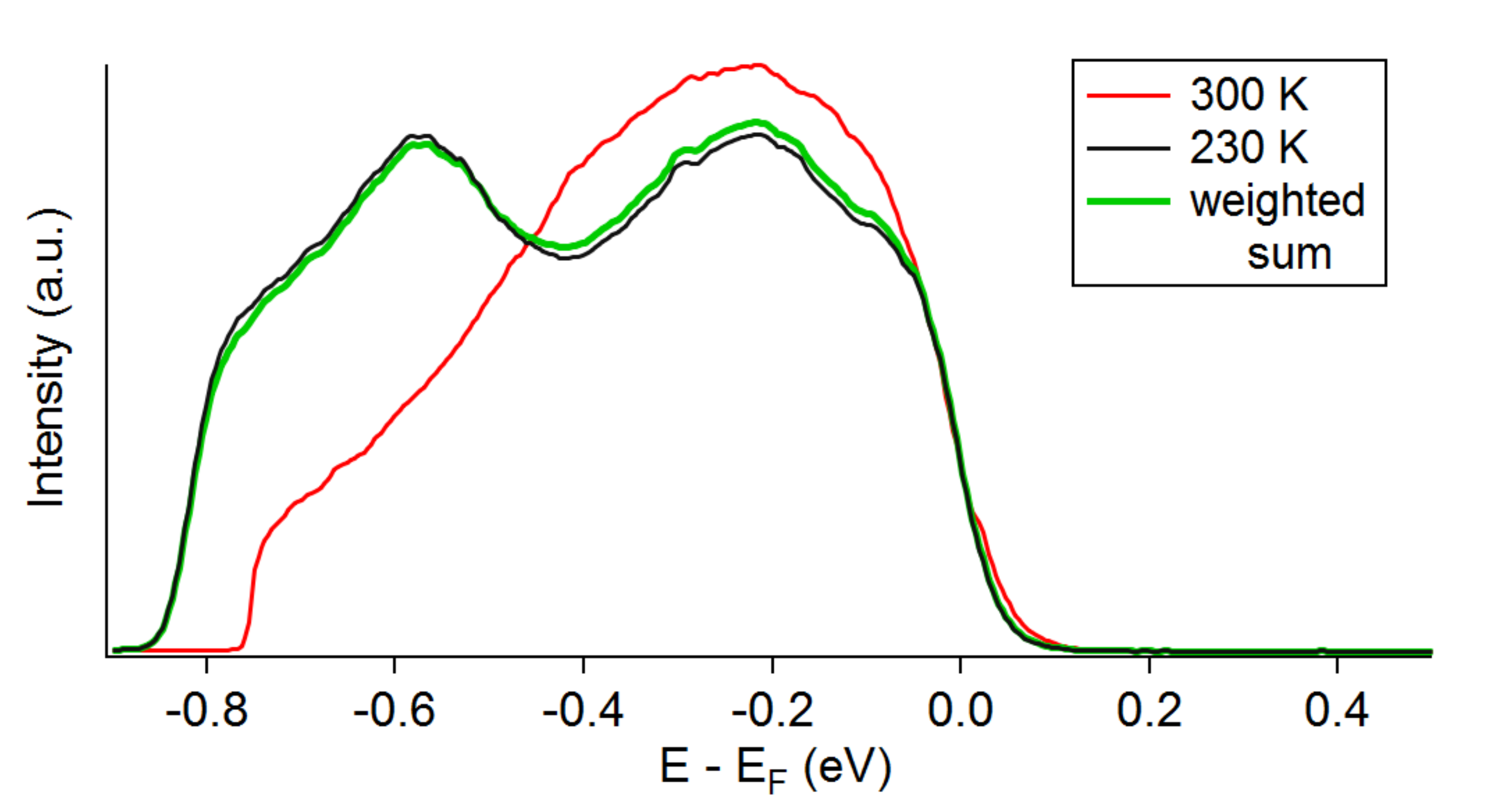}
\end{center}
\caption{\label{fig_new}
Static (angle-integrated) EDCs taken at 300 K (in red, RT phase) and at 230 K (in black, LT1 phase). Their weighted sum (green thick line) is obtained as the addition of $5\%$ of the static RT spectrum to $95\%$ of the static LT1 spectrum.
}
\end{figure}

In Fig. \ref{fig_2}(d), we show the time-dependent photoemission intensity of excited electrons at about 0.4 eV above $E_F$ (integrated in box 1 on Fig. \ref{fig_2} (b)), together with the time-dependent intensity in the upper edge of the CO peak (integrated in box 2). These two curves show very different behaviors: the green curve (box 1) displays the typical population dynamics for excited electrons near the Fermi level, while the blue curve (box 2) follows a step-like increase, with a slower rise than the green curve. The energy position of box 2 was chosen to minimize effects of the transient changes in the Fermi-Dirac distribution, while integrating the small transient changes near the CO peak. For this reason together with its unusal time-dependence, we infer that the photoemission changes below $-0.3$ eV are mainly photoinduced changes of the spectral function of \irte, rather than changes in the Fermi-Dirac distribution. Furthermore, given that the CO peak is the spectral signature of the LT1 phase, we conclude that it must be due to a (partial) photoinduced phase transition (PIPT) from the LT1 phase to the RT phase. This conclusion is further supported by Fig. \ref{fig_new}: here we add about $5\%$ of the static RT photoemission spectrum to $95\%$ of the static LT1 spectrum (at 230 K), in order to visualize what would be the spectral signature of a partial PIPT. This simple construction simulates a situation for which only $5\%$ of the probed volume would transit into the RT phase under the action of the pump pulse, e.g. as a consequence of sample inhomogeneity. The spectrum resulting from this combination is shown in green in Fig. \ref{fig_new}. It compares well to the EDCs recorded for delays greater than 1000 fs when the new transient state is established like shown in Fig. \ref{fig_2} (c). In particular, it should be compared to the EDC of Fig. \ref{fig_2} (c) measured at 4500 fs, since the transient change of the CO peak in Fig. \ref{fig_2} (d) indicates that the PIPT effects remain over a long timescale of more than 4 ps. This establishes the observation of a partial PIPT from LT1 to RT in \irte\ upon excitation with strong 825 nm pump pulses. Additionnally, our data reveal a persistant shift of the leading edge near $E_F$ of about 4 meV (see the inset in Fig. \ref{fig_2} (c)). This might be due to a transient change of the spectral function just above $E_F$, related to the band situated at about 0.12 eV (see red arrow in Fig. \ref{fig_2} (c)).

\begin{figure}
\begin{center}
\includegraphics[width=8.5cm]{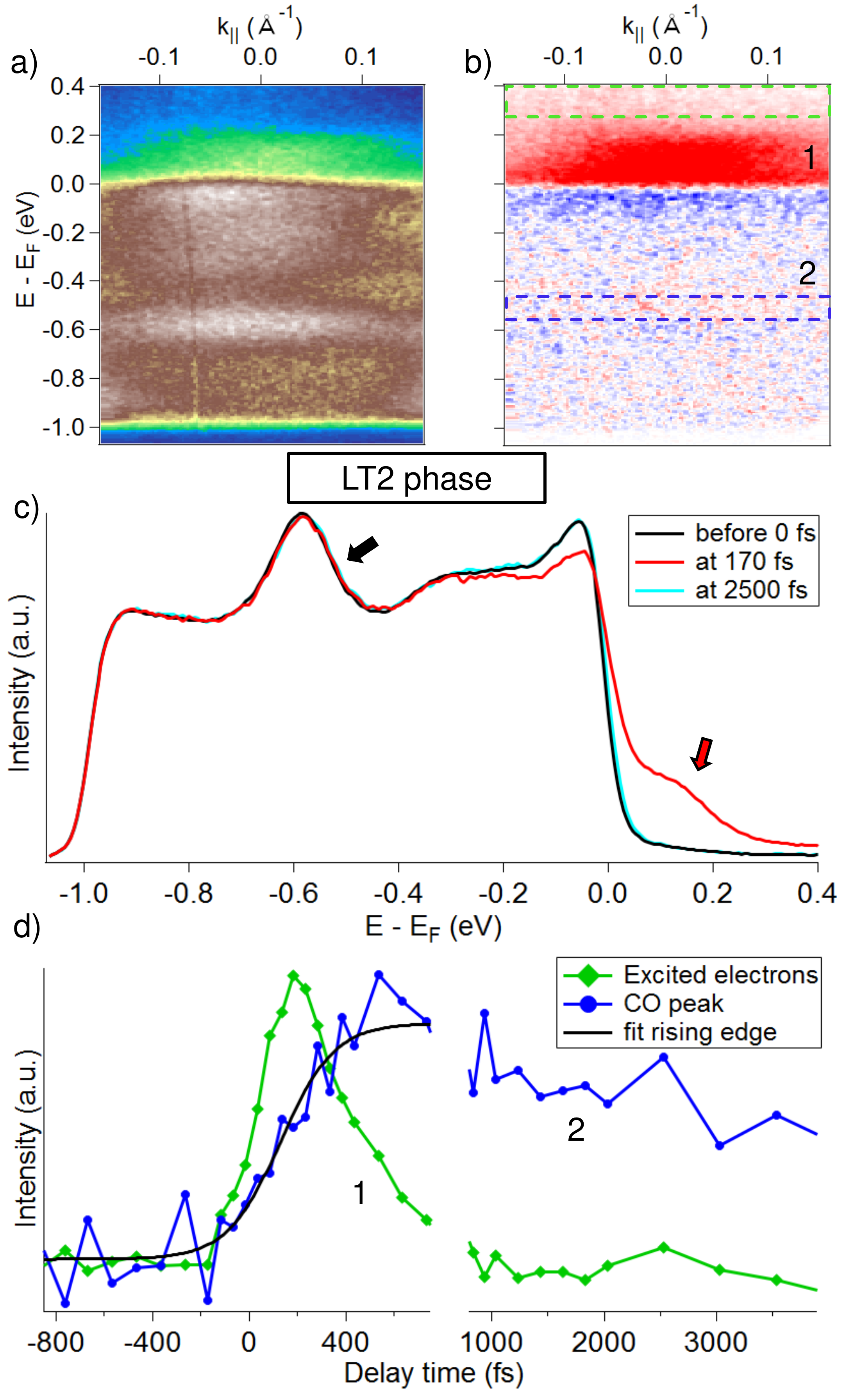}
\end{center}
\caption{\label{fig_3}
Time-resolved ARPES data in LT2 phase near $\bar{\Gamma}$, at 140 K. (a) Raw ARPES intensity map taken at + 170 fs and (b) the corresponding difference map (photoemission intensity difference between averaged data taken before 0 and data taken at 170 fs) for an absorbed fluence of 2.3 mJ/cm$^2$. (c) Corresponding (angle-integrated) EDCs at different delay times. (d) Transient intensities integrated in the box 1 (excited electrons) and 2 (CO peak) of graph (b), together with a step-like fit.
}
\end{figure}
In Fig. \ref{fig_3}, we show time-resolved ARPES data obtained in the LT2 phase. These data were acquired at 140 K with an absorbed excitation fluence of 2.3 mJ/cm$^2$. Fig. \ref{fig_3} (a) shows the raw data at 170 fs and Fig. \ref{fig_3} (b) the corresponding difference map. In Fig. \ref{fig_3} (c), EDCs are displayed before 0 fs, at 170 fs and at 2500 fs. The photoinduced changes observed here are similar to what has been observed in LT1 (see Fig. \ref{fig_2}), namely excited electrons above $E_F$ in a previously unoccupied band (see red arrow in Fig. \ref{fig_3} (c)), excited holes just below $E_F$ and photoinduced changes of the spectral function at higher binding energies, close to the CO peak (see black arrow in Fig. \ref{fig_3} (c)). In Fig. \ref{fig_3} (d), we show the time-dependent photoemission intensity integrated in the boxes 1 and 2 of Fig. \ref{fig_3} (b), following the dynamics of the excited electrons (green curve) in comparison to the spectral changes of the CO peak (blue curve) monitoring the PIPT. Similarly to the case of LT1, the dynamics of the PIPT are slower than the dynamics of the excited electrons. The CO peak intensity change, after having reached its maximum, relaxes back to its initial value on a very long timescale far above 3 ps.
We now apply a similar reasoning as used to construct Fig. \ref{fig_new} in order to figure out to which phases the PIPT occurs when starting from LT2 phase. From Fig. \ref{fig_1} (i), we know that the \irte\ spectral function changes very little between the LT1 and LT2 phases. Therefore mixing $5\%$ of the LT1 spectral function to $95\%$ of the LT2 spectral function would result in a spectrum that is hardly different from the LT2 phase spectrum and such a difference would not be visible in photoemission. We therefore conclude that the transient photoinduced changes measured in the LT2 phase essentially originate from a small volume fraction transiting to the RT phase.

\begin{figure}
\begin{center}
\includegraphics[width=8.5cm]{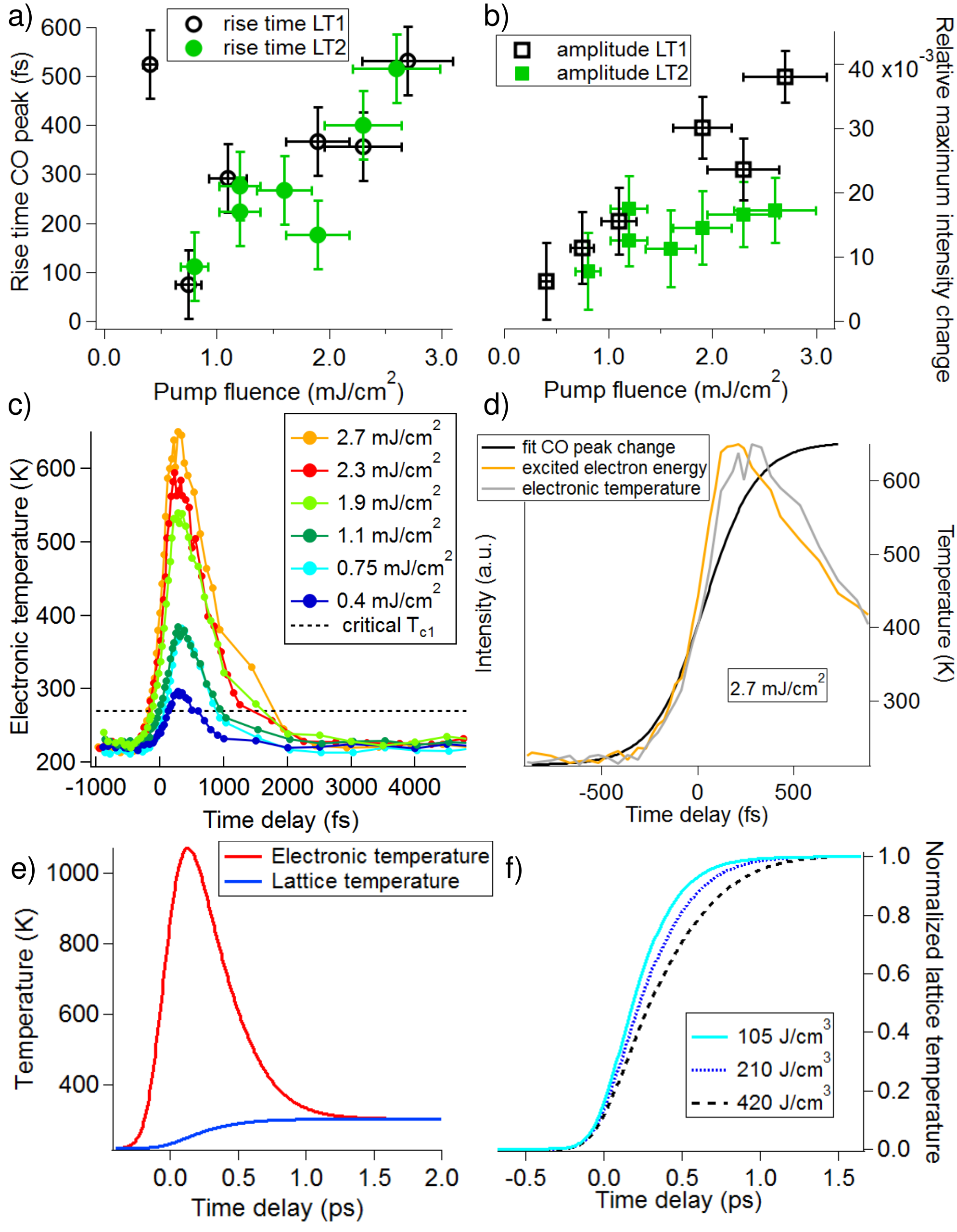}
\end{center}
\caption{\label{fig_4}
(a) Rise time of the PIPT obtained from fits to the photoemission intensity changes of the CO peak in LT1 and LT2 phases, as a function of absorbed pump fluence. (b) Maximum intensity change of the CO peak (relative to the maximum CO peak intensity) as a function of absorbed fluence. (c) Transient electronic temperature for different absorbed pump fluences in LT1. (d) Comparison between the transient mean excited electronic energy, electronic temperature and CO peak intensity in LT1 for an absorbed fluence of 2.7~mJ/cm$^2$. (e) Calculated transient electronic and lattice temperature for an absorbed energy density of 210 J/cm$^3$ using a two temperature model (see text for details). (f) Calculated transient lattice temperatures for three different absorbed energy densities and normalized to their maximum.
}
\end{figure}

In order to investigate the partial PIPT occuring in \irte, we have performed time-resolved ARPES for different excitation fluences in both LT1 and LT2 phases. For each fluence, we have fitted the time-dependent photoemission intensity of the CO peak with a broadened step function to extract its rise-time and maximum intensity changes as shown by black curves in Fig. \ref{fig_2} (d) and in Fig. \ref{fig_3} (d). In Fig. \ref{fig_4} (a) and (b), we plot the rise times and maximum intensity changes (relative to the maximum CO peak intensity) as a function of fluence, respectively. Globally, we see that the rise time and maximum intensity change increase with fluence. At the highest absorbed fluence achieved here (about 2.7 mJ/cm$^2$), the rise time, i.e. the time necessary to achieve the partial PIPT, reaches a duration as long as 500 fs, both in LT1 and LT2 phases. This is clearly a very slow process on the timescale of electron motion. Surprisingly, in Fig. \ref{fig_4} (a), the rise time of the lowest fluence used here (about 0.4 mJ/cm$^2$) for LT1 phase, is completely out of the trend and shoots up to 500 fs. Despite the low intensity of the corresponding photoinduced changes, this behavior is observed in the raw data (not shown here). To understand this exception, we analyze the slope of the electronic distributions just below $E_F$ for all data obtained in LT1 phase and extract an electronic temperature from fits of the Fermi-Dirac distribution between about -0.05 and 0.05 eV. The extracted values are shown in Fig. \ref{fig_4} (c) and we see that the maximum of each curve scales with the absorbed fluence. Interestingly, the electronic temperatures obtained for the lowest fluence are significantly lower than for the other fluences and hardly go over $T_{c1}$. This might indicate that different dynamics is at play here, because a threshold energy cannot be reached at this fluence.
We compare in Fig. \ref{fig_4} (d) the early evolution of the electronic temperature and the dynamics of the CO peak intensity change and of the mean excited electron energy $\left\langle E\right\rangle$. Here $\left\langle E\right\rangle$ is calculated by integrating the energy of excited electrons, $E-E_F$, weigthed with their photoemission intensity, $I(t)-I(t<0)$. In Fig. \ref{fig_4} (d), we see that the photoexcitation energy is quickly distributed into the excited electrons as observed within the detector window chosen here. The electronic temperature rises more slowly and reaches its maximum about 70 fs later. The main observation is that the increase in the intensity change of the CO peak is even slower and reaches its maximum at even longer times.

\section{Discussion}

The partial PIPT evidenced in this work involves small changes of the spectral function, which reach only a few percents of the intensity of the CO peak at the highest fluence used here (Fig. \ref{fig_4} (b)). This indicates that the LT phases of \irte\ are very robust against photoexcitation, at least as far as the electronic structure is concerned. It is tempting to emphasize the similarity with the PIPT observed in VO$_2$, for which incident fluences as high as 25 mJ/cm$^2$ were necessary to achieve the complete PIPT in a material having an absorption length of 180 nm at 800 nm wavelength\cite{WallPRB,BerglundVO2}. In time-resolved ARPES studies \cite{WegkampVO2,YoshidaPRB2014}, photoinduced changes of only a few percents have been obtained for incident fluences above 6 mJ/cm$^2$. In VO$_2$, the structural phase transition is of first order like in \irte. However, the similarity stops here, since the PIPT has been shown to occur on a very fast timescale\cite{CavalleriPRB} in VO$_2$, being as fast as 80 fs. In the case of \irte, we observe a slow (partial) PIPT, with durations of the order of several hundreds of fs. Furthermore, we see that this PIPT rise time (see Fig. \ref{fig_4} (a)) is \textit{increasing} for \textit{increasing} fluences.
This tells us that the time necessary to achieve the complete spectral changes observed here is dictated by a slow process governed by the pump pulse energy flow. We propose naturally that it is due to the energy flow from the excited electrons into the lattice, as also indicated by the hierarchy of timescales seen in Fig. \ref{fig_4} (d): first the pump energy is absorbed by the electrons, which are excited more than 1 eV above $E_F$. These electrons then scatter down towards $E_F$ and thermalize, and the maximum electronic temperature is reached only $\sim 100$ fs later. We conjecture that this remarkable delay is due to the presence of a presumably electron-like band in the unoccupied states, about 0.12~eV above $E_F$ (see red arrow in Fig. \ref{fig_2} (c) and Fig. \ref{fig_3} (c)). A central observation here is that the dynamical spectral signature of the PIPT is even slower than the electronic temperature. This is again compatible with our proposition that the PIPT is driven by the transient lattice temperature, which rises more slowly than the electronic temperature.

The fluence dependence of the rise time of the CO peak change (Fig. \ref{fig_4} (a)) can be understood qualitatively with a simple two temperature model analysis, including the transient electronic and lattice temperatures. For this purpose, we compute the electronic and lattice temperatures for \irte, according to the following differential equations
\begin{eqnarray}
\frac{\partial T_e}{\partial t} &=& -\gamma(T_e)\cdot(T_e-T_l)+\frac{P}{C_e},\nonumber\\
\frac{\partial T_l}{\partial t} &=& \frac{C_e}{C_l}\gamma(T_e)\cdot(T_e-T_l).\nonumber
\end{eqnarray}
The coupling function, $\gamma(T)=3\lambda\Omega_0^2/(\hbar^2\pi T)$, is inversely proportional to temperature\cite{Allen}. The differential equations can be derived from the model used by Perfetti \textit{et al.}\cite{PerfettiBISCO} but for only one phonon subsystem.
We take the electronic specific heat $C_e$ as linear in the electronic temperature $T_e$ and a temperature dependent lattice specific heat $C_l$ fitted to experimental values\cite{Fang2013}. The latent heat at the phase transition at $T_{c1}$ is much smaller than the energy density deposited by the pump pulse (see below), and is therefore neglected.
From the optical data of Fang \textit{et al.}\cite{Fang2013,WangPC}, the absorption length in \irte\ at 825 nm is about 20 nm, which means that 1 mJ/cm$^2$ of absorbed fluence corresponds to an absorbed energy of 210 J/cm$^3$. In our example here\footnote{We also use a phonon energy of 30 meV, an electron-phonon coupling constant $\lambda$ of 0.2, and the linear term of the electronic specific heat is $1.5\cdot10^{-4}J/(\text{cm}^3\text{K}^2)$. However the choice of these parameters does not influence our qualitative argument.}, we first use an absorbed energy density of 210 J/cm$^3$. The calculated temperatures are shown in Fig. \ref{fig_4} (e) and are typical for the output of the two temperature model: the electronic temperature rises as fast as the pump pulse intensity and reaches very high temperatures, while the lattice temperature rises more slowly due to the finite energy exchange rate between the electrons and the lattice and attains a lower temperature, 300 K in our example. On Fig. \ref{fig_4} (f), the calculated lattice temperatures are shown for three different absorbed energy densities and normalized to their maximum. One sees that the higher the absorbed energy density, the more time it takes for the lattice temperature to reach $95\%$ of its maximum. 
This behavior is due to the complex temperature dependencies in the two temperature model and can be mainly traced back to the coupling function $\gamma(T)\propto 1/T$, which indicates that for increasing electronic temperatures, the energy transfer from electrons to lattice becomes less efficient.
As a consequence, the timescale of the CO peak change, which represents the transient lattice temperature, increases with fluence (Fig. \ref{fig_4} (a)) because for higher fluences (or absorbed energy densities), it will take more time for the CO peak change to reach its maximum transient value. In this interpretation, the PIPT timescale is given by the transient lattice temperature and can occur only as fast as the transient heating of the lattice.

The photoinduced spectral changes after a few ps do not exactly coincide with the simple superposition of static ARPES data shown in Fig. \ref{fig_new}. 
In particular, we see significant photoinduced spectral changes on the peak at -0.2 eV (see e.g. Fig. \ref{fig_2}(c)), in addition to a persistent shift of the leading edge at $E_F$ and no modification of the sample work function, which changes drastically through the phase transition in the static data. 
It is also very surprising that we can induce only small changes of the spectral function in our experiment, given the high fluences used here and the short absorption length of \irte. 
As written above, 1 mJ/cm$^2$ of absorbed fluence corresponds to an absorbed energy of 210 J/cm$^3$.
This represents a large energy density, which should result in a transient electronic temperature raising up to more than 1000 K and subsequently a lattice temperature raising above 300 K. From this estimation, and even from the observation that electronic temperatures of several 100s K above RT are reached at $E_F$ (Fig. \ref{fig_4}(c)), it is clear that the excitation of electrons in \irte\ is not sufficient to trigger efficiently the PIPT.
The latent heat of the first order structural phase transition (about 20 J/cm$^3$, from Ref. \onlinecite{KoNatureComm}) cannot explain this fact, since it is very small compared to the absorbed energy densities used here.
This might be an indication that the system is in fact trapped in a metastable out-of-equilibrium state between LT1 and RT phases. It is therefore interesting to recall that the equilibrium phase transition taking place in \irte\ involves both a change of its unit cell symmetry \textit{and} of the charge order, which breaks the translational symmetry. In a recent paper, Ivashko \textit{et al.} have shown that, upon pressure, these two transitions occur at different temperatures for samples with a few percents of Pt doping\cite{IvashkoIrTe}. We propose here that a metastable out-of-equilibrium state involving only one of these two structural transitions could be achieved in \irte\ upon photoexcitation. This is further supported by the fact that no transient change of the sample work function is observed here.
We hope that this will trigger interest for further time-resolved studies.

It has been proposed that the first-order CO phase transitions taking place in \irte\ are due to a Mott phase transition involving spin-orbit coupled states\cite{KoNatureComm}. The phenomenology of successive first-order Mott phase transitions in \irte\ is reminiscent of the case of TaS$_2$\cite{SiposTaS2,RossnagelReview}. TaS$_2$ has been intensively studied by time-resolved techniques and it has been shown that the Mott gap at $\Gamma$ collapses quasi-instantaneously upon photoexcitation, what has been understood as evidence for its electronic origin\cite{PetersenTaS2,PerfettiTaS2,HellmanNatComm}. Interpreting the CO peak in \irte\ as the lower Hubbard band of a Mott system, one would expect its response to photoexcitation to be ultrafast, well below the time-resolution of our experiment. The small shift of the CO peak observed in our study, which could be viewed as the precursor of a Mott gap collapse, is by far too slow to be related to the ultrafast physics of photoexcited Mott insulators. Our time-resolved ARPES study of \irte\ does therefore not give any evidence for Mott physics in this material.

\section{Conclusion}

In conclusion, we have studied the ultrafast dynamics of the first-order structural phase transitions in \irte\ with time-resolved ARPES. We observe that a partial phase transition can be photoinduced in this material using strong near-infrared pump pulses. The time necessary for this PIPT is increasing for increasing pump fluence and eventually becoming as high as 500 fs. Furthermore, this characteristic time is slower than the timescales necessary for depositing energy in the electronic subsystem and raising its temperature. We deduce that the partial PIPT is driven by the energy transfer from excited electrons into the lattice. However, the observed photoinduced changes in ARPES are small despite the large absorbed energy. We conjecture that this material is actually trapped in an out-of-equilibrium state.
From the above we conclude that both the phase transitions in thermal equilibrium and those following ultrafast excitation are mainly driven by a lattice instability rather than by a Mott instability.

\section{Acknowledgements}
Part of this work was performed at the PEARL beamline of the Swiss Light Source of the Paul Scherrer Institut in Switzerland. We thank J. Osterwalder for fruitful discussions. C.M. acknowledges the support by the SNSF grant No. $PZ00P2\_ 154867$. We are grateful to Cephise Cacho for sharing the time-resolved ARPES analysis software. This work was supported by the Swiss National Science Foundation through Div. II. F.W. was supported by the Helmholtz Society under contract VH-NG-840. M.M. was supported by the Karlsruhe Nano-Micro Facility (KNMF).

\end{document}